\documentclass[sigconf, authorversion=true, nonacm=true]{acmart}

\copyrightyear{2022}
\acmYear{2022}
\setcopyright{acmcopyright}
\acmConference[NANOCOM '22]{The Ninth Annual ACM International Conference on Nanoscale Computing and Communication}{October 5--7, 2022}{Barcelona, Spain}
\acmBooktitle{The Ninth Annual ACM International Conference on Nanoscale Computing and Communication (NANOCOM '22), October 5--7, 2022, Barcelona, Spain}
\acmPrice{15.00}
\acmDOI{10.1145/3558583.3558875}
\acmISBN{978-1-4503-9867-1/22/10}

\def\BibTeX{{\rm B\kern-.05em{\sc i\kern-.025em b}\kern-.08emT\kern-.1667em\lower.7ex\hbox{E}\kern-.125emX}}

\usepackage{caption}
\usepackage{subcaption}
\usepackage{nicefrac}
\usepackage{siunitx}
\usepackage{array,framed}
\usepackage{booktabs}
\usepackage{
  color,
  soul,
  float,
  epsfig,
  wrapfig,
  graphicx
}
\usepackage{setspace}
\usepackage{latexsym,fancyhdr,url}
\usepackage{enumerate}
\usepackage{algorithm2e}
\usepackage{algpseudocode}
\usepackage{graphics}
\usepackage{xparse} 
\usepackage{xspace}
\usepackage{multirow}
\usepackage{csvsimple}
\usepackage{balance}

\usepackage{
  tikz,
  pgfplots,
  pgfplotstable
}
\usepackage{hyperref}

\usetikzlibrary{
  shapes.geometric,
  arrows,
  external,
  pgfplots.groupplots,
  matrix
}

\pgfplotsset{compat=1.9}

\usepackage{mathtools}

\DeclareMathAlphabet{\mathcal}{OMS}{cmsy}{m}{n}

\DeclareGraphicsExtensions{%
    .png,.PNG,%
    .pdf,.PDF,%
    .jpg,.mps,.jpeg,.jbig2,.jb2,.JPG,.JPEG,.JBIG2,.JB2}

\usepackage{xparse}
\newcommand{\bnm}{\begin{newmath}}
\newcommand{\enm}{\end{newmath}}

\newcommand{\bea}{\begin{eqnarray*}}%
\newcommand{\eea}{\end{eqnarray*}}%

\newcommand{\bne}{\begin{newequation}}
\newcommand{\ene}{\end{newequation}}

\newcommand{\bal}{\begin{newalign}}
\newcommand{\eal}{\end{newalign}}

\newenvironment{newalign}{\begin{align}%
\setlength{\abovedisplayskip}{4pt}%
\setlength{\belowdisplayskip}{4pt}%
\setlength{\abovedisplayshortskip}{6pt}%
\setlength{\belowdisplayshortskip}{6pt} }{\end{align}}

\newenvironment{newmath}{\begin{displaymath}%
\setlength{\abovedisplayskip}{4pt}%
\setlength{\belowdisplayskip}{4pt}%
\setlength{\abovedisplayshortskip}{6pt}%
\setlength{\belowdisplayshortskip}{6pt} }{\end{displaymath}}

\newenvironment{newequation}{\begin{equation}%
\setlength{\abovedisplayskip}{4pt}%
\setlength{\belowdisplayskip}{4pt}%
\setlength{\abovedisplayshortskip}{6pt}%
\setlength{\belowdisplayshortskip}{6pt} }{\end{equation}}

\newcounter{ctr}

\newcounter{mytable}
\def\mytable{\begin{centering}\refstepcounter{mytable}}
\def\endmytable{\end{centering}}

\newcounter{myfig}
\def\myfig{\begin{centering}\refstepcounter{myfig}}
\def\endmyfig{\end{centering}}

\newlength{\saveparindent}
\newlength{\saveparskip}
\newcommand{\E}{{\rm I\kern-.3em E}}

\newcommand{\figref}[1]{\mbox{Figure~\ref{#1}}}

\renewcommand{\eqref}[1]{\mbox{Equation~(\ref{#1})}}

\newcommand{\tabref}[1]{\mbox{Table~\ref{#1}}}

\def \part {part}

\renewcommand{\paragraph}[1]{\vspace*{6pt}\noindent\textbf{#1}\;}

\def \blackslug{\hbox{\hskip 1pt \vrule width 4pt height 8pt
    depth 1.5pt \hskip 1pt}}
\def \qed{\quad\blackslug\lower 8.5pt\null\par}

\newcounter{mynote}[section]

\newcommand\ignore[1]{}

\newcounter{rcnote}[section]

\newcounter{mrnote}[section]

\newcounter{fknote}[section]

\newcounter{anote}[section]

\DeclareMathSymbol{\mlq}{\mathord}{operators}{``}
\DeclareMathSymbol{\mrq}{\mathord}{operators}{`'}

\newcommand{\rhf}[2]{R_{f, \gamma}}

\DeclareDocumentCommand{\edist}{o o}{
  \ensuremath{
    \IfNoValueTF{#1}{{d}}{{\sf d}(#1,#2)}
  }
}

\newcommand{\olrk}[1]{\ifx\nursymbol#1\else\!\!\mskip4.5mu plus 0.5mu\left(\mskip0.5mu plus0.5mu #1\mskip1.5mu plus0.5mu \right)\fi}

\NewDocumentCommand{\indseq}{ O{1} O{r} }{{#1}\ldots {#2}}

\begin{document}
\title{Towards Spatial Multiplexing in Wireless Networks within Computing Packages}

\author{F\'atima Rodr\'iguez-Gal\'an}
\affiliation{%
 \institution{Universitat Polit\`{e}cnica de Catalunya}
  \city{Barcelona}
  \country{Spain}
  }
\email{fatima.yolanda.rodriguez@upc.edu}

\author{Elana Pereira de Santana}
\affiliation{%
  \institution{University of Siegen}
 \city{Siegen}
 \country{Germany}}
\email{Elana.PSantana@uni-siegen.de}

\author{Peter Haring Bol\'ivar}
\affiliation{%
  \institution{University of Siegen}
 \city{Siegen}
 \country{Germany}}
\email{peter.haring@uni-siegen.de}

\author{Sergi Abadal}
\affiliation{%
 \institution{Universitat Polit\`{e}cnica de Catalunya}
  \city{Barcelona}
  \country{Spain}
  }
  \email{abadal@ac.upc.edu}

\author{Eduard Alarc\'on}
\affiliation{%
 \institution{Universitat Polit\`{e}cnica de Catalunya}
  \city{Barcelona}
  \country{Spain}
  }
  \email{eduard.alarcon@upc.edu}

\renewcommand{\shortauthors}{Rodríguez-Gal\'an et al.}

\begin{abstract}
Wireless Networks-on-Chip (WNoCs) are regarded as a disruptive alternative to conventional interconnection networks at the chip scale, yet limited by the relatively low aggregate bandwidth of such wireless networks. Hence, any method to increase the amount of concurrent channels in this scenario is of high value. In this direction, and since WNoC implies close integration of multiple antennas on a chip anyway, in this paper we present a feasibility study of compact monopole antenna arrays in a flip-chip environment at millimeter-wave and sub-terahertz frequencies. By means of a full-wave solver, we evaluate the feasibility to create, at will, concentrations of field in different spots of the chip. This way, we set the steps towards spatial multiplexing that enables concurrent multicast communications and also increases the aggregate bandwidth of the wireless network. Our results at 60 GHz show two clearly separable parallel channels that radiate simultaneously from two opposite corners of the chip, achieving a Signal-to-Interference Ratio (SIR) of around 40 dB, which proves that the channels are independent of each other even in such an enclosed environment. Further, we see potential to expand our approach to three or more concurrent channels, and to frequencies beyond 100 GHz.

\end{abstract}



\maketitle

\section{Introduction}
\label{sec:intro}
The Network-on-Chip (NoC) paradigm and, more recently, its analogous Network-in-Package (NiP) have become  the de facto standard for the interconnection of cores in multicore processors. However, as we enter the hyperscaling era \cite{salahuddin2018era}, the  communication requirements increase up to a point where conventional NoCs and NiPs alone  may  not  suffice. Their limited  scalability  is  in fact  turning  communication  into  the  performance  bottleneck of  manycore  systems,  thus  calling  for  new  solutions  at  the interconnect level \cite{bertozzi2015fast, ganguly2022interconnects}.

\begin{figure}
   \centering
   \includegraphics[width=\columnwidth]{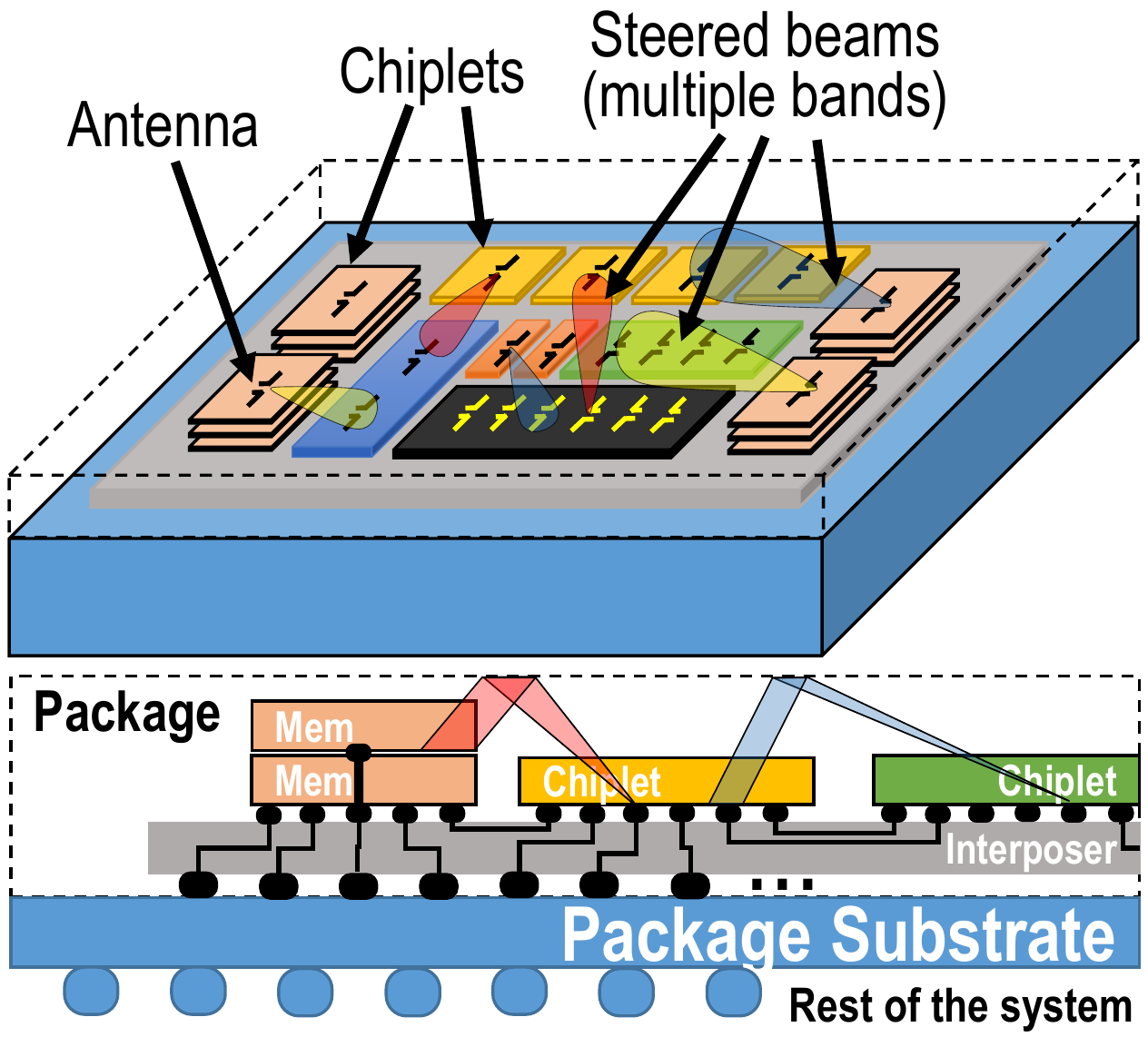}
 	\caption{Schematic diagram of a System-in-Package (SiP) hosting a heterogeneous set of chiplets. The interconnect fabric is composed of a silicon interposer Network-in-Package (NiP) augmented, as proposed in this paper, with multiple concurrent spatial wireless links.}
   \label{fig:vision}
  \vspace{-0.1cm}
 \end{figure}

Advances in integrated  antennas  and  transceivers have  led  to  the  proposal of  Wireless  Network-on-Chip  (WNoC)  as  a  complement  of  or  alternative  to  existing NoCs.  WNoCs  basically  consist of  the  co-integration  of  RF  front-ends  with  cores  or  clusters of cores \cite{Laha2015}.
Radio  waves  propagate  through  the package at nearly the speed of light until reaching the intended destinations, also located within the same package, as shown in Figure \ref{fig:vision}. At the receivers, signals are then demodulated and  deserialized \cite{timoneda2020engineer}.  Since  intermediate  router hops are avoided,  WNoC  reduces  the  latency  of  long-range and broadcast communications by an order of magnitude. On the  downside,  wireless  bandwidth is  limited  and  needs  to  be shared among the cores.
As a result, Medium Access Control (MAC) protocols or multiplexing methods are  required  to  avoid  collisions  and  interference  in the  WNoC. However,  this  approach  has  important limitations because the number of non-overlapping frequency, code,  or  time  slotted  channels  achievable  in  this  resource-constrained scenario is relatively small \cite{abadal2018medium, Kodi2015}. A large number of channels increases the complexity of hardware in frequency multiplexing or synchronization components in code multiplexing.

An alternative or complement to the multiplexing schemes mentioned above would be spatial multiplexing as proposed in the authors' prior work \cite{Obeamforming} and pictorially represented in Figure \ref{fig:vision}. Although the creation of multiple concurrent spatial channels comes with a loss of the broadcast nature of wireless communications, which is very much appreciated by computer architects \cite{franques2021widir}, there are several applications that require multicast and can still benefit from having multiple spatial channels \cite{abadal2016characterization}. In any case, such an approach requires the use of antennas with a relatively large aperture, which may be complex to integrate with on-chip environments. 

Indeed, to implement the on-chip array, the  flip-chip  package  does  not  leave  much  space  for the antennas. Due  to  the presence  of  solder  bumps,  antennas cannot be implemented in the first metal layer \cite{Zhang2019}. Alternatively, designers  have  to  use  the  metal  layers  closer to  the silicon, where patch antennas and printed dipoles might be implemented. However, the proximity to the virtual ground plane formed by the array of micro-bumps reduces their efficiency, whereas co-planarity between antennas further increases losses; also, as we are in heavily area constrained scenario one must be cautious with the use of available space \cite{timoneda2018channel}. Finally, as  proposed  recently, one could fabricate  Through-Silicon Via (TSV) to implement vertical monopoles \cite{timoneda2018channel, Pano2020a}, an approach followed by us. Due to the stringent area constraints of the scenario, directional antennas and MIMO arrays are generally prohibitive \cite{timoneda2020engineer}.

Even so, for the evaluation of feasibility of MIMO channels and near field communications, a statistical characterization of a 2$\times$2 MIMO wireless link operated in a mode-stirred cavity is carried out in \cite{lodro2021statistical}. The authors obtained two parallel channels within the confined space for different conditions of the mode-stirred enclosure. They provide a characterization of a complex MIMO channel in highly reflective propagation environments. 

In \cite{Phang2018}, the possibility of improving the performance of wireless RF interconnect in the near field is studied. The paper presents a design to get multi-channel communications between antennas in their mutual near field. This approach is useful for applications such as on-chip and chip-to-chip wireless communications.

Further, to increase communication distance and capacity in the THz band, the concept of Ultra-Massive MIMO is introduced in \cite{Akyildiz2016}. The authors propose the design of graphene-based plasmonic nano-antenna arrays composed of thousands of elements comprised in a few square millimiters. They contemplate reception, transmission, beamforming, spatial multiplexing and multi-band communication schemes as applications for the resulting array. Although the previously mentioned papers provide points to consider in our work, the proposed arrays and MIMO characterizations done by them do not precisely occur in a chip environment. 

In \cite{Baniya2018a, Baniya2018}, a four element array for beam switching in chip-to-chip communications in a multi-chip system at 60 GHz is proposed. Pattern changes are achieved by switching the elements on and off. However, their package scenario is different than the one we use. Narde \emph{et al.} \cite{Narde2019, Narde2020} evaluates the beamforming and transmission capabilities of on-chip arrays for intra-chip and inter-chip communications in multi-chip systems. They use a four zigzag antenna element phased array at 60GHz to study static beamsteering in specific directions. In their simulations, they arranged the four elements with a spacing of 0.75 mm which might imply problems with the use of available space in the chip. Moreover, the array is not reconfigurable. The previously cited papers mainly contemplate planar antennas in a chip environment different from the one we use and that we will describe in the next section.

In this paper, we attempt to create spatial channels or field distributions concentrated in different parts of the chip, by building 4$\times$4 monopole antenna arrays in a flip-chip package and exploring different phase distributions. Hence, our strategy relies on having reconfiguration abilities at the transmitter and receiver antennas, which is in contrast to recent proposals using near-passive programmable reflectarrays at the boundaries of the environment to somehow re-structure the reverberant behavior of the channel (an idea proposed for indoor communications \cite{DelHougne2019} and recently ported to the on-chip domain \cite{imani2021smart}). 

In our proposed approach, we take advantage of the high permittivity of the bulk silicon layer used in the chip to create compact arrays with vertical monopoles. Also, the silicon high losses prove to be useful to attenuate waves that can interfere in adjacent channels and to place antennas close together without much coupling. Towards that goal, we conducted simulations to assess the appropriate distance among the antennas of the array and achieved compact 4$\times$4 arrays at 60 GHz approximately 1 mm\textsuperscript{2} and less than -10 dB coupling among adjacent antennas. Further, with the adjusting of excitation phases, we got two parallel channels, which we demonstrated along two different directions within the computing package. The antenna arrays radiate simultaneously with little interference with each other, achieving a very large Signal-to-Interference Ratio (SIR) of around 40 dB. Moreover, the results seem to be extendable to more channels and more frequency bands, as we show. 

The remainder of this paper is organized as follows. In Section \ref{sec:background}, we describe the simulation methodology and the antennas used in this work. In Section \ref{sec:methodology}, we explain the criteria followed to design of the array and show the results obtained via full-wave simulations. Section \ref{sec:evaluation_channel} outlines the post-processing steps carried out to evaluate spatial multiplexing. In Section \ref{sec:conclusion}, the paper is concluded.








\section{Background}
\label{sec:background}
Here, we first depict the simulation environment in Section \ref{sec:env}, to then describe the steps followed to design the vertical monopole that serves as the basis of our arrays in Section \ref{sec:antenna}.

\subsection{Environment Description}
\label{sec:env}
\begin{figure}[!t]
\centering
\includegraphics[width=\columnwidth]{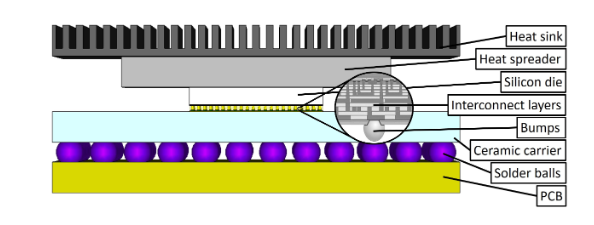}
\vspace{-0.6cm}
\caption{Schematic of the layers of a flip-chip package.}
\label{fig:flipchip}
\vspace{0.2cm}
\end{figure}

The environment structure is a flip-chip package as shown in \figref{fig:flipchip}.  
In this configuration, the chips are turned over and connected to the system substrate through a set of solder bumps. The packaged chip therefore has the silicon substrate on top, which is in turn interfaced by the spreader material and system heat sink on top.   The insulator and metal stack are placed at the bottom, interfaced by the solder bumps that connect it to the system \cite{timoneda2018channel, Elmasri2019}.

\begin{table}[!t] 
\caption{Package parameters.} 
\label{tab:flipParams}  
\vspace{-0.1cm}
\centering
\begin{tabular}{ccccccc} 
\hline
{\bf Parameter} & {\bf Thickness} & {\bf Materials} & {\bf Units} \\
\hline
Heat Spreader & 0.5 & Aluminum Nitride & mm \\
Silicon die & 0.5 & Bulk Silicon & mm \\
Lateral Space & 0.5 & Vacuum & mm\\
Chiplet insulator & 0.01 & SiO2 & mm \\
Bumps & 0.0875  & Copper &  mm\\
Frequency & 60 & -& GHz \\
\hline
\end{tabular}
\end{table}

The layers are described from top to bottom as summarized in \tabref{tab:flipParams}. On top the heat spreader, modeled as Aluminum Nitride, dissipates the heat out of the silicon since it has good thermal conductivity; with $\varepsilon_{r}$=8.6 and $\rho$=0.0003. The insulator is silicon dioxide with $\varepsilon_{r}$=3.9 and $\rho$=0.025. 
The silicon die, made of bulk silicon, serves as the foundation of the transistors. This layer has $\varepsilon_{r}$=11.9 and low resistivity (10 $\Omega\cdot$cm), which is convenient for the operation of transistors, but not for electromagnetic propagation since it attenuates the signal. Finally, we simulate the interconnects and the bumps as one solid layer of copper \cite{timoneda2018channel}. Our chip has a size of 10x10 mm\textsuperscript{2} and is to be studied at a frequency of 60 GHz in a full-wave solver.  
The solver allows us to perform phase sweeps on the excitation of each antenna of the array, obtain the S-parameters to assess the coupling between antennas and to study the field distribution for those phase changes. Our choice of solver for this work is CST Microwave Studio \cite{CST}.

\subsection{Single Antenna Design}
\label{sec:antenna}
Due to its good \emph{a priori} lateral coupling and opportunistic compatibility with conventional chip pacakge designs, we have chosen a vertical monopole antenna as baseline for our study. The monopole antenna is modeled as a thin and long cylindrical metallic structure, placed vertically passing through the silicon and fed from the first metal layers. Practically, this can be implemented by fabricating TSVs that emerge from the metallization layers and prematurely stopping the fabrication before reaching the heat spreader. Since the bumps layer is seen as a solid metallic block of metal at 60 GHz, due to the small bump pitch, this layer acts as a sort of ground plane for the monopole, increasing the effective antenna length due to image theory \cite{timoneda2018channel}. Hence, we initially set the monopole length $L$ to
\begin{equation} \label{equ}
L = \frac{\lambda}{4} = \frac{v_ {p}}{4\cdot f} = \frac{c_0}{4\cdot \sqrt{\varepsilon_{Si}}\cdot f}
\end{equation} 
where $c_0$ is the speed of light, $f = 60$ GHz is the target frequency, and $\varepsilon_{Si}$ is the permittivity of silicon in that frequency region.

\begin{figure}[!t]
\centering
\includegraphics[width=\columnwidth]{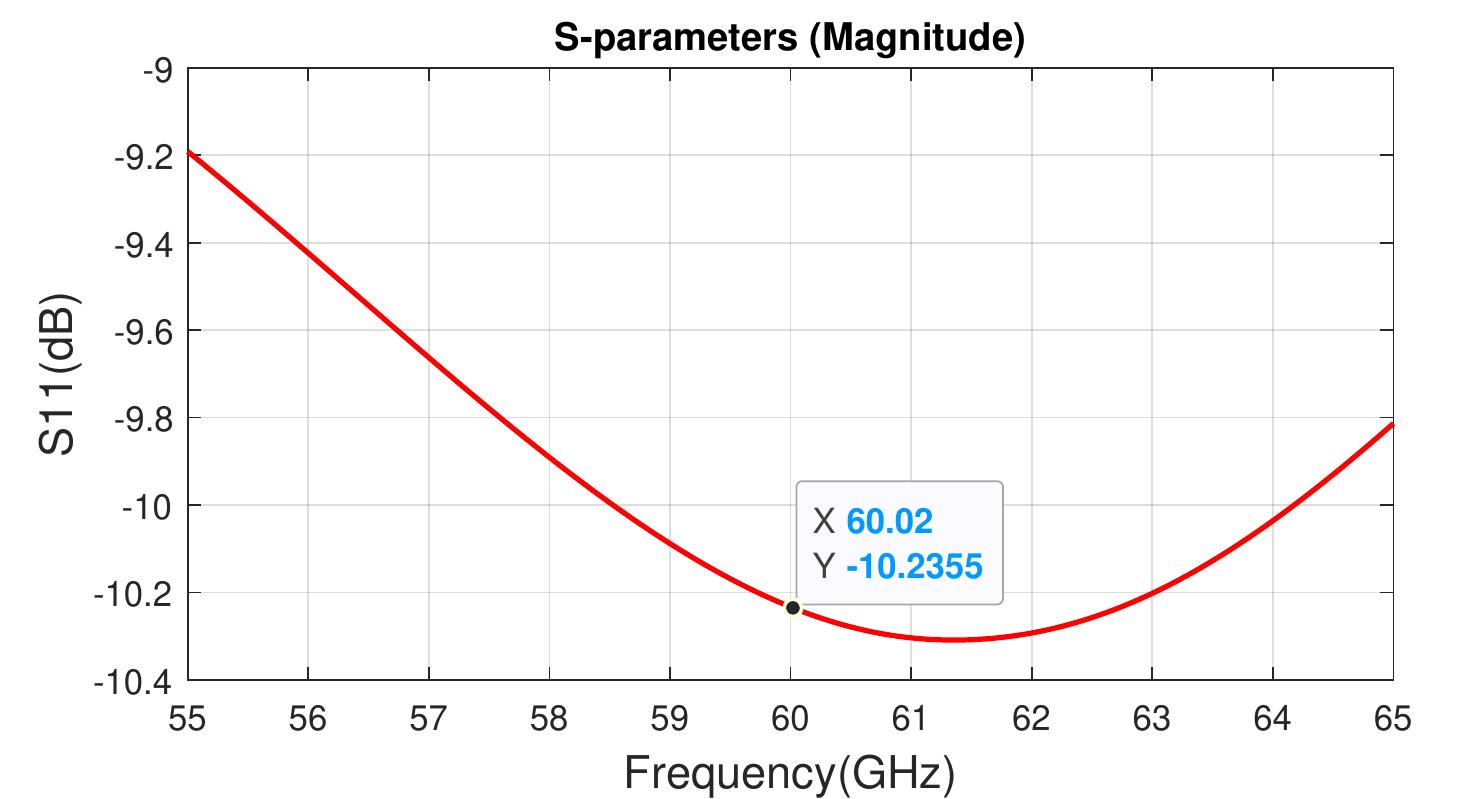}
\vspace{-0.1cm}
\caption{S11 of the monopole within a flip-chip environment.} 
\label{fig:S11_one_antenna}
\end{figure}

While the monopole fits entirely within the silicon layer, its proximity to the interface with the heat spreader material with a different permittivity may lead to a shift in the resonance. Taking this into account, to adjust the dimensions of our antenna we first model a simple scenario with a quarter-wave monopole sized using \eqref{equ}. Afterwards we introduced the monopole in the chip environment and we fine-tuned the length to get a good reflection coefficient close to 60 GHz.~\figref{fig:S11_one_antenna} shows the reflection coefficient of the monopole 
working at 60 GHz in the flip-chip environment.

\begin{figure}[!t]
\centering
\includegraphics[width=.8\columnwidth]{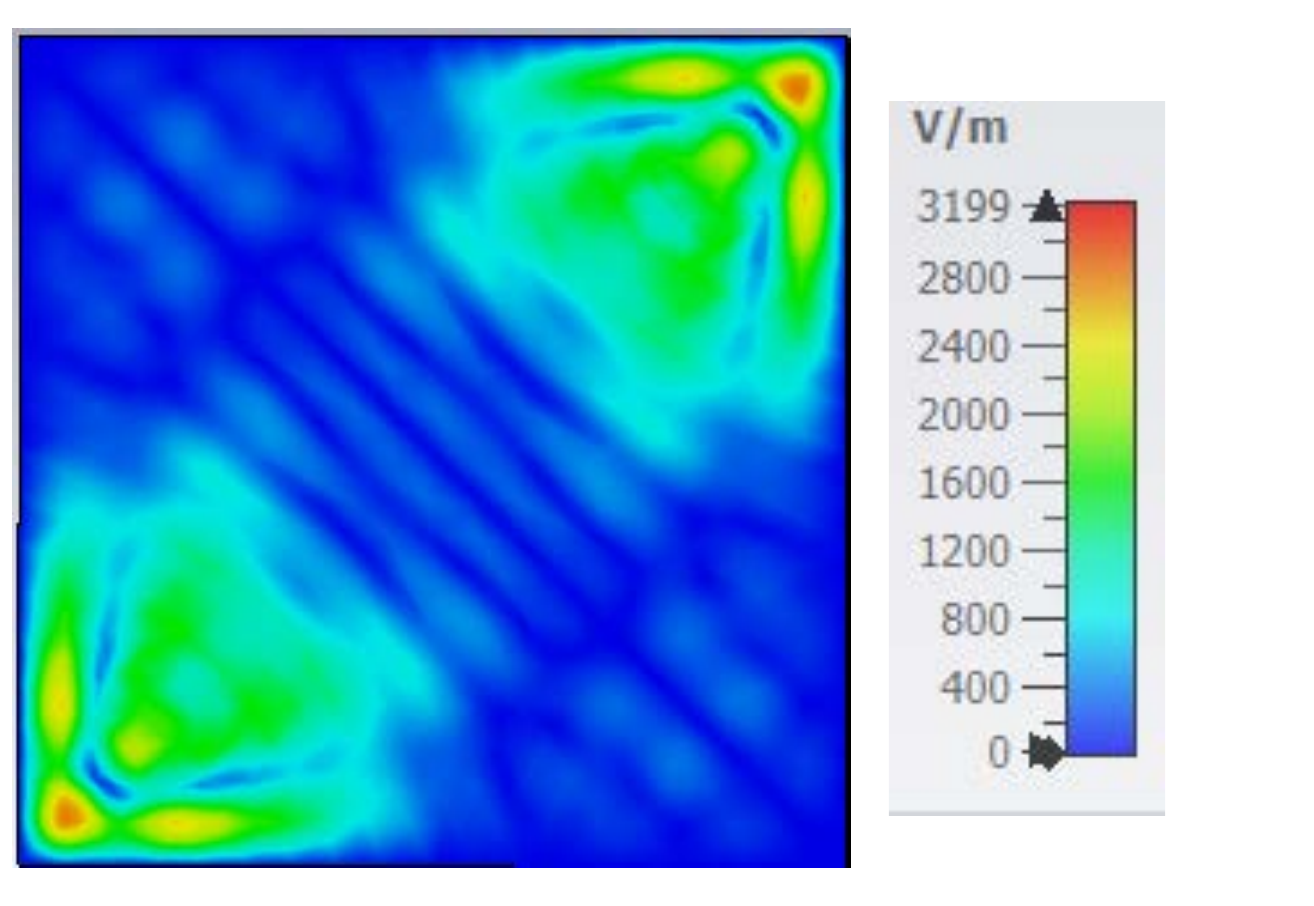}\hspace{-0.2cm}
\caption{Electric field generated by two monopole antennas placed in opposite corners and transmitting at the same time.}
\label{fig:Two_monopoles}
\end{figure}

\figref{fig:Two_monopoles} shows two monopoles placed diagonally on the chip and radiating at the same time. Some phase sweeps were made to the antennas excitation to attempt to create a sort of \emph{beam} or field concentration in any part of the chip. However, as expected, none of the sweeps yield any controllability or appreciable differences in the field concentration for that matter. In fact, the beam-like distribution pointing to the center of the chip is due to the proximity of vertical walls at the lateral limits of the chip. Therefore, to be able to direct the beam or create concentration of energy in any part of the chip, we will need to use antenna arrays.

\section{Antenna Arrays within a Computing Package}
\label{sec:methodology}

\subsection{Mutual Coupling}

To get a better grasp of the behavior of the antennas in the flip-chip package, the scenario was simplified. Tests were made with only upper and lower boundaries of the chip, without any lateral vacuum spacers. We also experimented with distance between antennas, number of elements and their position on the chip to get to our final design. We will be using monopoles with a length of 0.475 mm and a radius of 0.005 mm, as discussed in Section \ref{sec:background}.

Although the bulk silicon permittivity allows us to create a compact array, we need to make sure that the distance between elements will not lead to losses due to mutual coupling. Such a coupling issue was studied placing two monopoles in the corner of the chip forming a small array, then exciting them with the same phase at the same time, and finally monitoring the S-parameters to evaluate the coupling in each case. We simulate $\lambda$/20, $\lambda$/10, $\lambda$/8, $\lambda$/5, $\lambda$/4 and $\lambda$/2 where $\lambda = \tfrac{c_0}{\sqrt{\varepsilon_{Si}}\cdot f}$ is the wavelength in silicon. 

\begin{figure}[!t]
\includegraphics[width=\columnwidth]{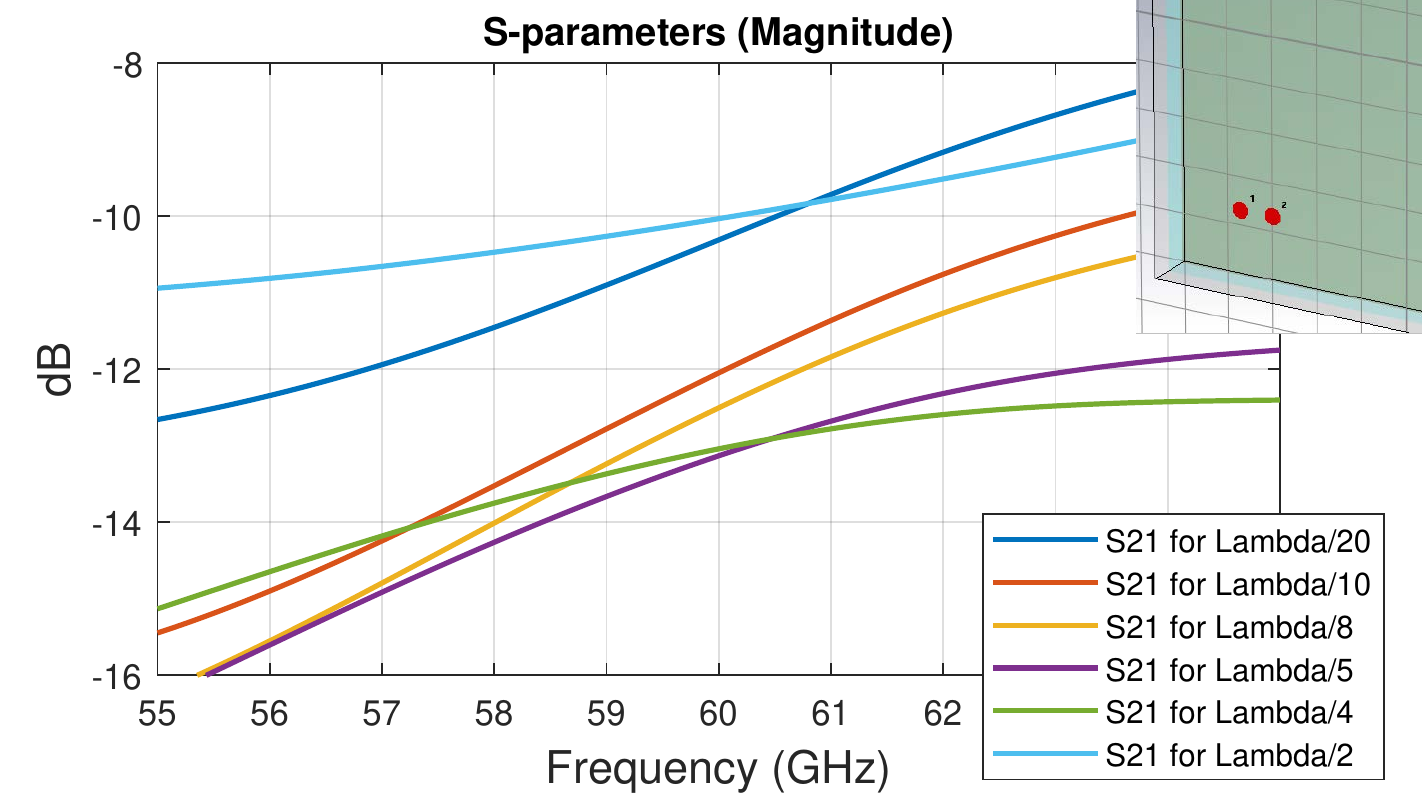}
\caption{Landscape used for the coupling assessment and results at different distances between two antennas.}
\label{fig:S21}
\end{figure}

\begin{figure}[!t]
\includegraphics[width=\columnwidth]{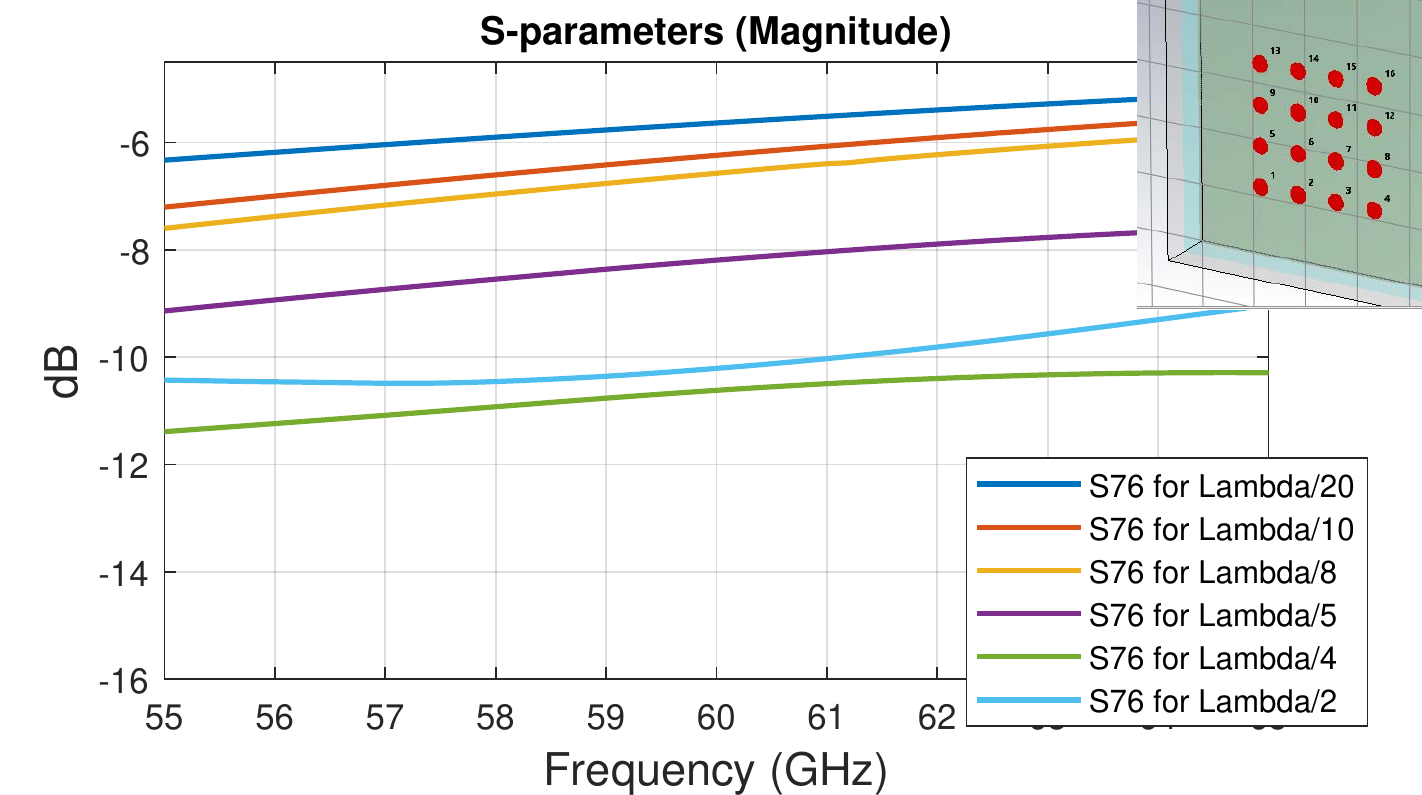}
\caption{Landscape of the array and test coupling results for different distances among the elements.} 
\label{fig:S21_array}
\end{figure}

\figref{fig:S21} presents the distance dependency for the S-parameters and demonstrates that even for minimum distances between elements the coupling seems to remain very low; so we should be able to form arrays with short distances among antennas. However, because we are aiming for an array, we test the coupling issue at the mentioned distances for a 16-element array configured in a 4$\times$4 manner. Then, we use two antennas in the center of the array to check the accumulated influence of the surrounding elements. In our configuration, these were antennas 6 and 7. 
In \figref{fig:S21_array} we observe the harmful effects of adding more antennas and lowering the distance among them, leading to an inter-element coupling worse than -10 dB when the distance is 
smaller than $\lambda/4$.

In the tests carried out, when using distances like $\lambda$/2, we obtained interesting field concentrations plots. However, as we are working on the chip scale, one must be cautious with the use of available space. Also, some phase combinations that prove to create a clear \emph{beam} for smaller distances, at $\lambda$/2 start to lose shape. For the smallest distances, namely $\lambda$/20, $\lambda$/10, $\lambda$/8, although we can create and direct the beam using a correct combination of excitation phases, we face with the undesired coupling issue. For $\lambda$/20, in addition to the coupling effects, fabricating TSVs with the required pitch is difficult and expensive. With these results, the best compromise is to explore the energy concentration with a 16-element array with distances of $\lambda$/4 among each other.
As a result, the evaluated array occupies a chip area of approximately 1.18 mm\textsuperscript{2}; which represents 1.18\% of the 100 mm\textsuperscript{2} chip and even less in larger dies.

\subsection{Field Distribution}
Finding a combination of phases for the array that give us a clear beam and certain controllability is complex in this scenario. Instead of using an analytic approach we use the post-processing combine results tool offered by CST, to get changes in the energy patterns. 
This means that after we run the simulation of our environment, we applied a macro that made a sweep of excitation phases re-using and combining the fields provided by the solver's field monitor at 60 GHz. This method allows us to obtain more results in less time, to experiment with different phase changes dynamically and without much computational cost. Such an approximation is valid since the coupling among the different elements is below -10 dB, as we have shown in Figure \ref{fig:S21_array}.
Also, we made a smaller environment with one monopole. Using array theory and the array factor tool offered by CST, we manipulated the radiation pattern of the antenna creating a virtual array and the best results were used to compose the beams in the flip chip scenario. 
We gradually increased the difficulty of the simulations. First stimulating a single element and adding more as we became acquainted with the changes that certain phases caused in its field pattern until we reached the array of 16 elements.

\begin{figure}[!t]
  \includegraphics[width=1\columnwidth]{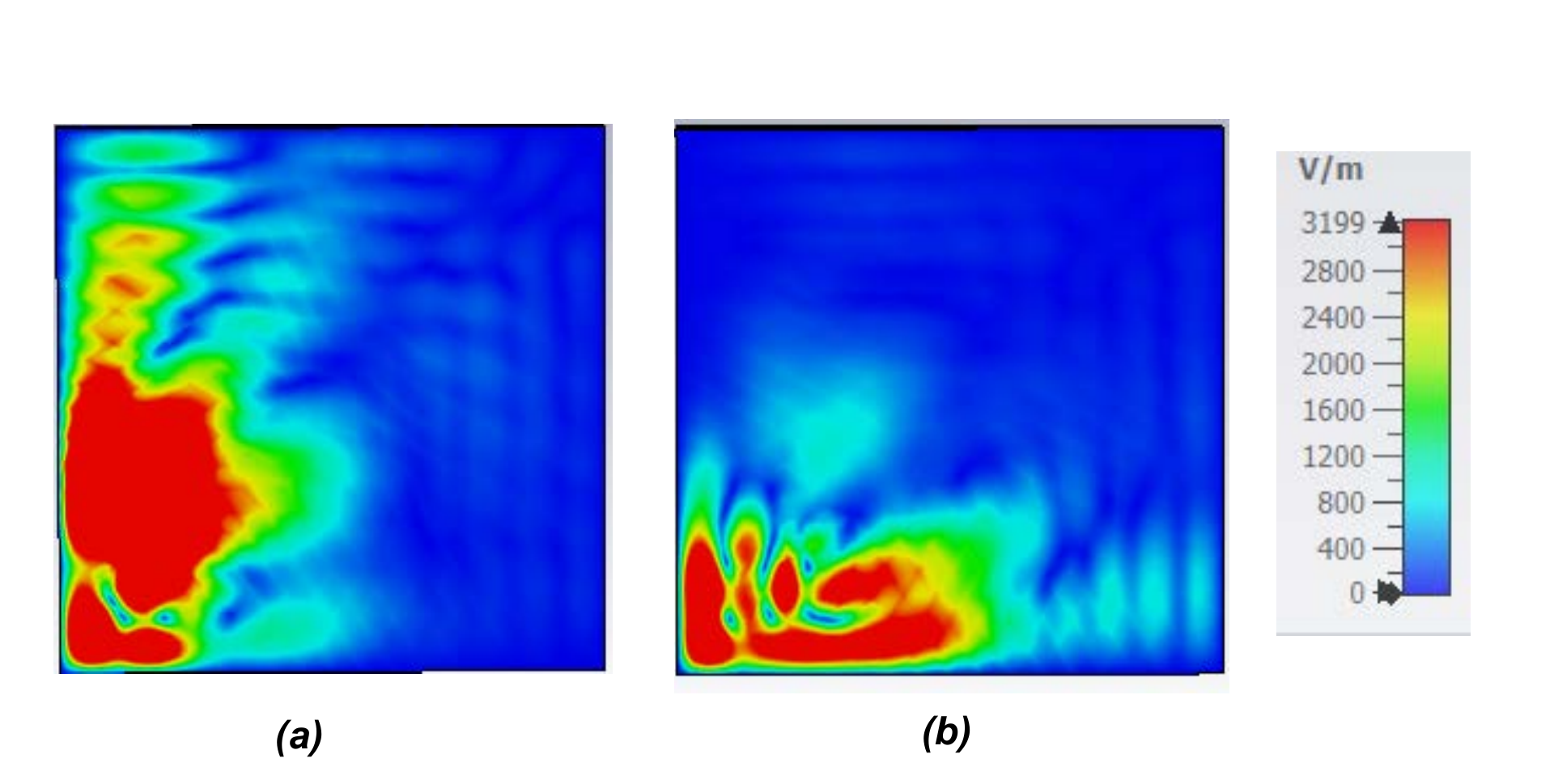}\hfill
  \vspace{-0.2cm}
 \caption{Field distributions of a phased array with configurations to steer the field along (a) the Y axis and (b) the X axis of the coordinate system.}
  \label{v_h_beams}
\end{figure}

\tabref{table_vertical} shows a summary of the final excitation phases used on each antenna to obtain parallel channels. In \figref{v_h_beams}, we see the results of such a phase profile combination. The 4$\times$4 array (array 1) is placed on the bottom left corner of the chip and radiates towards its opposite corners with a clear and well shaped beam. 


\begin{table}
\vspace{0.4cm}
 \caption{Phase values leading to the field shown in \figref{v_h_beams}.}
 \label{table_vertical}
 \vspace{-0.1cm}
 \begin{tabular}
{|l|c|c|c|c|c|c|c|c|}
\hline
\multicolumn{9}{|c|}{Vertical Beam} \\ \hline
Port & 1 & 2 & 3 & 4 & 5 & 6 & 7 & 8 \\ 
Phase & 90 & 120 & 150 & 180 & 0 & 30 & 60 & 90 \\ \hline
Port & 9 & 10 & 11 & 12 & 13 & 14 & 15 & 16 \\ 
Phase & -90 & -60 & -30 & 0 & -180 & -150 & -120 & -90 \\ \hline
\hline
\multicolumn{9}{|c|}{Horizontal Beam} \\ \hline
Port & 1 & 2 & 3 & 4 & 5 & 6 & 7 & 8 \\ 
Phase & 0 & -330 & -300 & -270 & -150 & -120 & -90 & -60 \\ \hline
Port & 9 & 10 & 11 & 12 & 13 & 14 & 15 & 16 \\ 
Phase & 60 & 90 & 120 & 150 & 270 & 300 & 330 & 0 \\ \hline
\end{tabular}
\vspace{-0.4cm}
\end{table}


\begin{figure}[!t]
  \includegraphics[width=1\columnwidth]{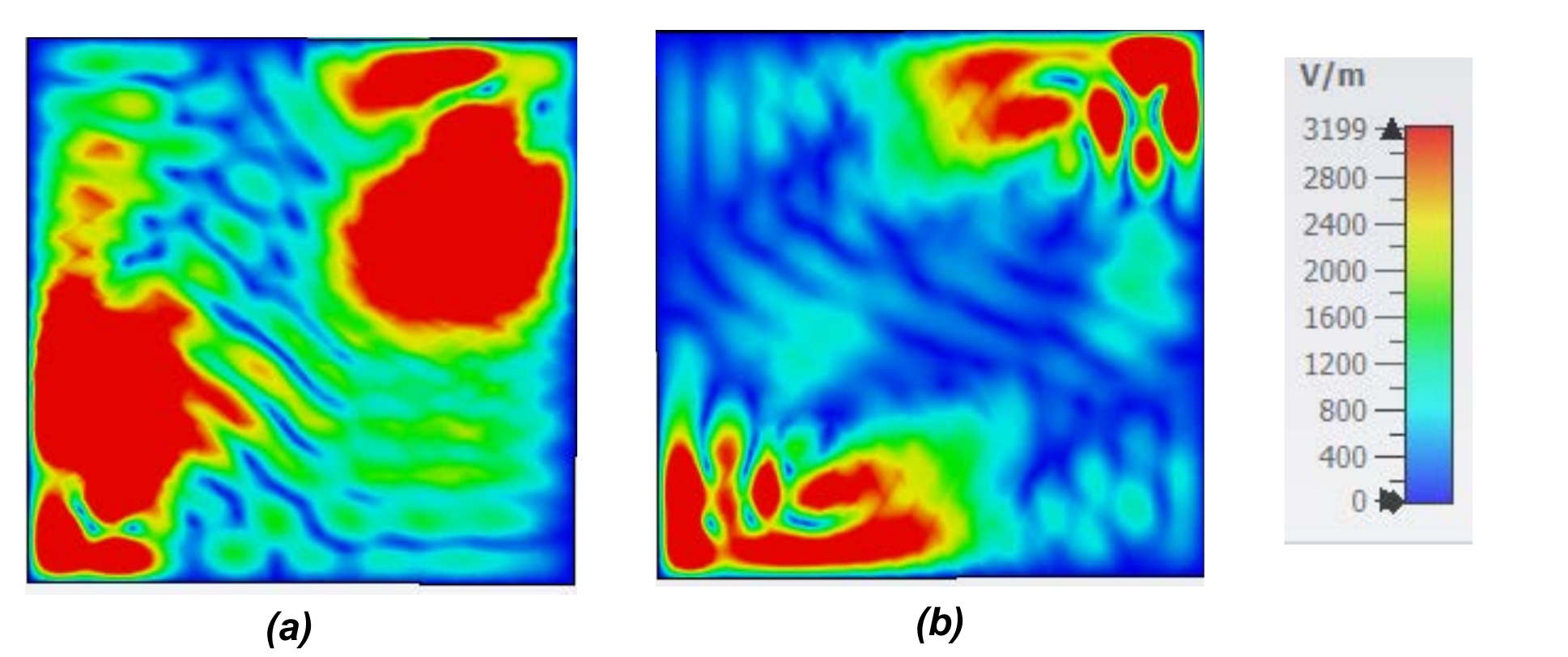}
  \vspace{-0.2cm}
 \caption{Field distributions of two phased arrays with configurations to steer the field towards the opposite corner along (a) the Y axis and (b) the X axis of the coordinate system.}
 \label{two_v_h_beams}
\end{figure}

Our next step is to come up with a combination that holds two parallel channels radiating at the same time without interfering with each other. To do this, we place an identical array in the upper right corner (array 2), we perform a phase sweep procedure as described above. This sweep is based on the results obtained with the array factor in the one monopole scenario.

\figref{two_v_h_beams} presents the results of our simulations. At first sight it seems that we manage to create two parallel concentrations of energy that radiate at the same time, in principle without interfering with each other. This already proves one of our main goals, which was to create beams inside a chip using antenna arrays.




\section{Evaluation of In-Package Spatial Channels}
\label{sec:evaluation_channel}
To verify that the channels obtained in the previous section radiate as independently as they appear to do, another post-processing step is performed by using the resulting fields of the phase manipulation. In Section \ref{sec:baseline}, we outline the method and assess our basline case of two channels at 60 GHz. Then, we show the results of scaling the system to 110 GHz in Section \ref{sec:freq} and of scaling the number of intended spatial channels to at least three in Section \ref{sec:channels}.

\subsection{Baseline: Two Spatial Channels at 60 GHz} \label{sec:baseline}
The post-processing made to further validate the beams along the Y axis is seen in ~\figref{fig:pp}. We take the field created when only the array on the bottom left corner (array 1) radiates and subtract from it the field produced when both arrays radiate. This gives us the level of interference on array 1 when array 2 radiates. From the image, it is observed that the space where array 1 dominates is clearly along its Y axis, with bright colors, whereas the other side of the chip clearly shows dominance from array 2. Overall, both channels are separated by way more than 20 dB of interference, hence they are isolated from each other. The SIR gives us a measure of the reliability of the channel in this case. From the image we see that the radiation from array 1 arrives to the intended opposite corner with a SIR of more than 40 dB, meaning that the interference level is very low when both arrays radiate simultaneously.

\subsection{Scaling the Frequency} \label{sec:freq}
\begin{figure*}[!t]
\begin{subfigure}[b]{0.5\textwidth}
    \includegraphics[width=0.48\textwidth]{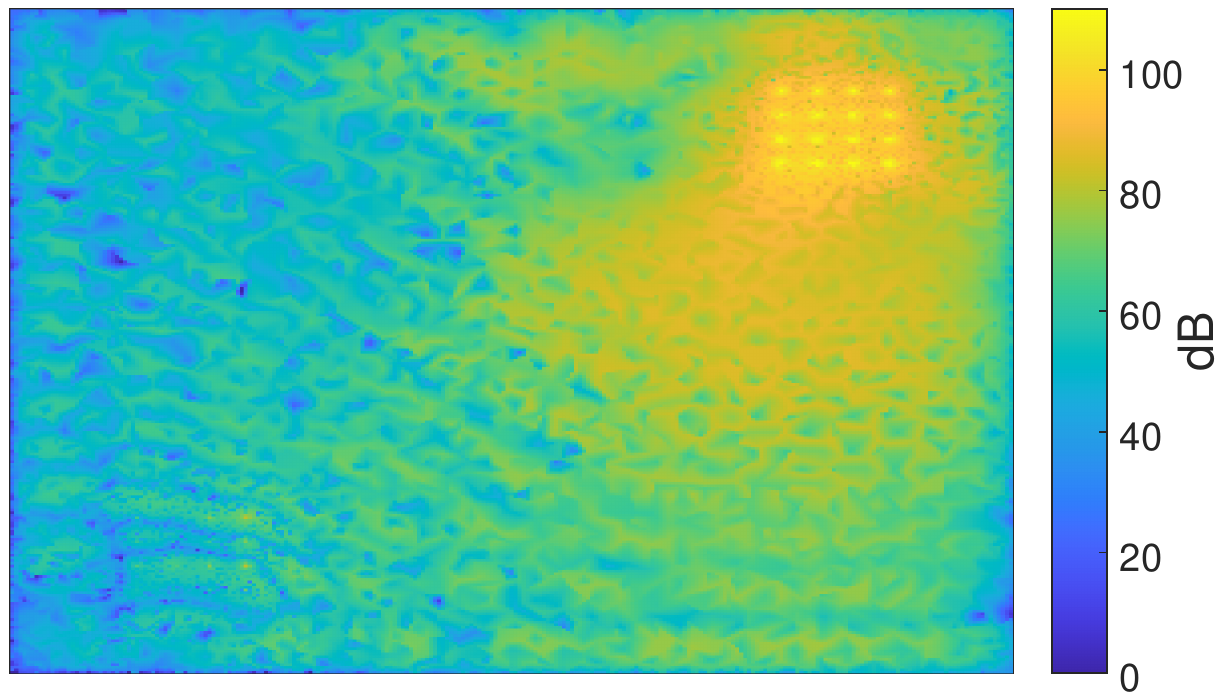}
    \includegraphics[width=0.48\textwidth]{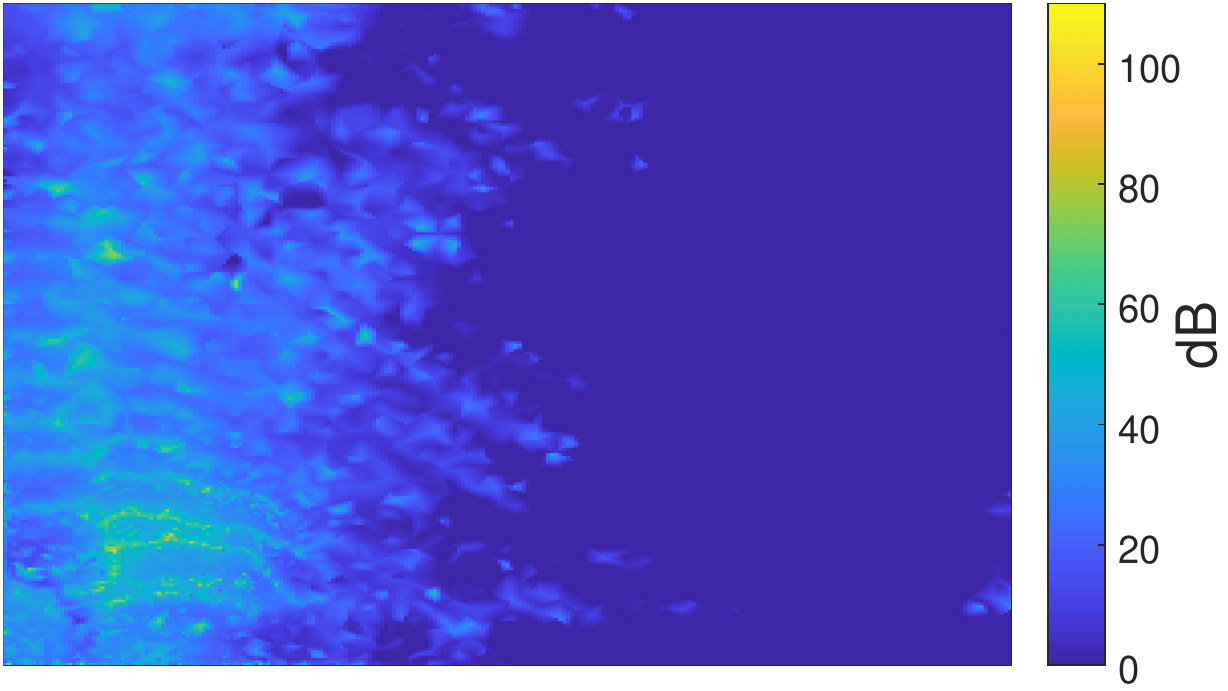}
\caption{Multiple channels along the X axis at 60 GHz.} 
    \label{fig:pp}
    \end{subfigure}
    \begin{subfigure}[b]{0.47\textwidth}
    \includegraphics[width=0.48\textwidth]{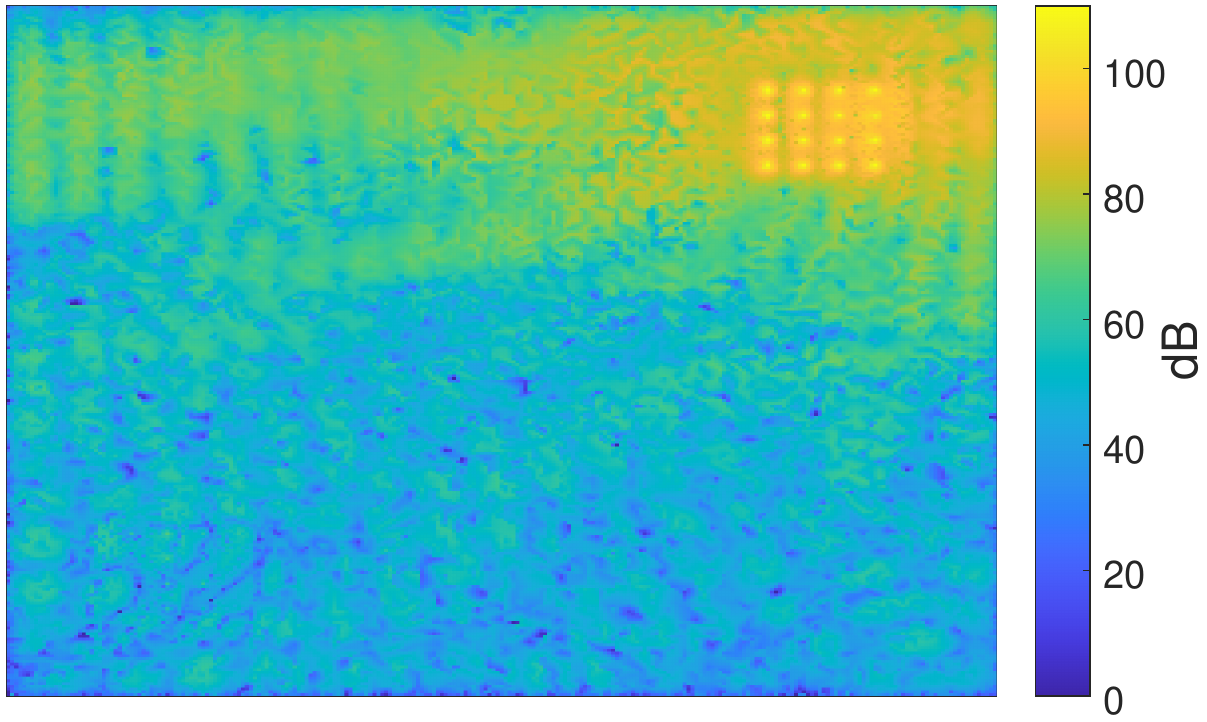}
\includegraphics[width=0.48\textwidth]{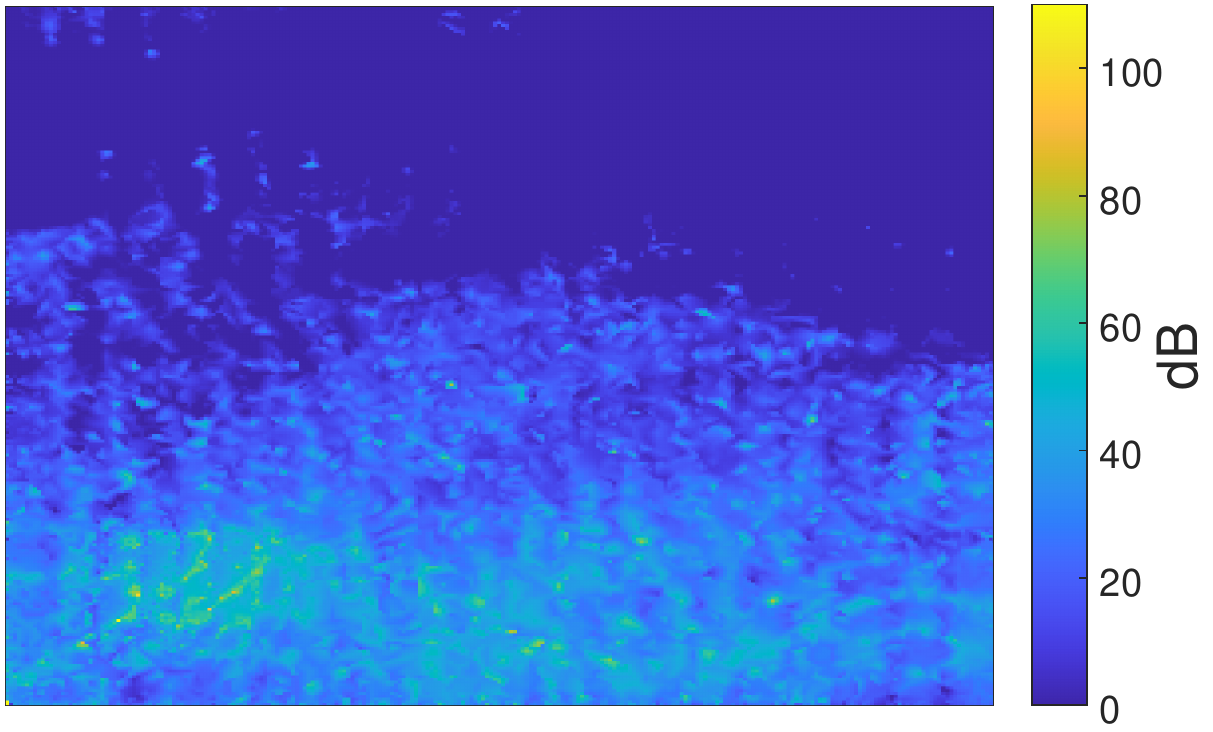}
\caption{Multiple channels along the Y axis at 110 GHz.}
\label{fig:110GHz}
\end{subfigure}
\vspace{0.2cm}
\caption{Interference field (left) and Signal-to-Interference Ratio (SIR, right) in two different cases.}
\end{figure*}

In the WNoC use case, it is desirable to achieve diversity both on frequency and space for multi-channel communications. For this, we contemplate to open spatial channels in frequencies other than 60 GHz. To evaluate this, we assume the same landscape and antennas at 60 GHz and simulate it at 110 GHz to consider that the monopole is working close to a harmonic of the original 60 GHz tone. In our simulations, the S11 parameter of the monopole drops to -10.3dB, but still it can be considered that monopole resonates so we can use it for our purposes. \figref{fig:110GHz} shows the interference field and the SIR relationship of the scenario at hand, but at 110 GHz. For this frequency, we also achieved two well-defined independent channels, hence opening the door to joint space-frequency multiplexing to maximize the channel creation within the package.

\begin{figure*}[!t]
\begin{subfigure}[b]{1.1\columnwidth}
 \centering
 \includegraphics[width=1.1\columnwidth]{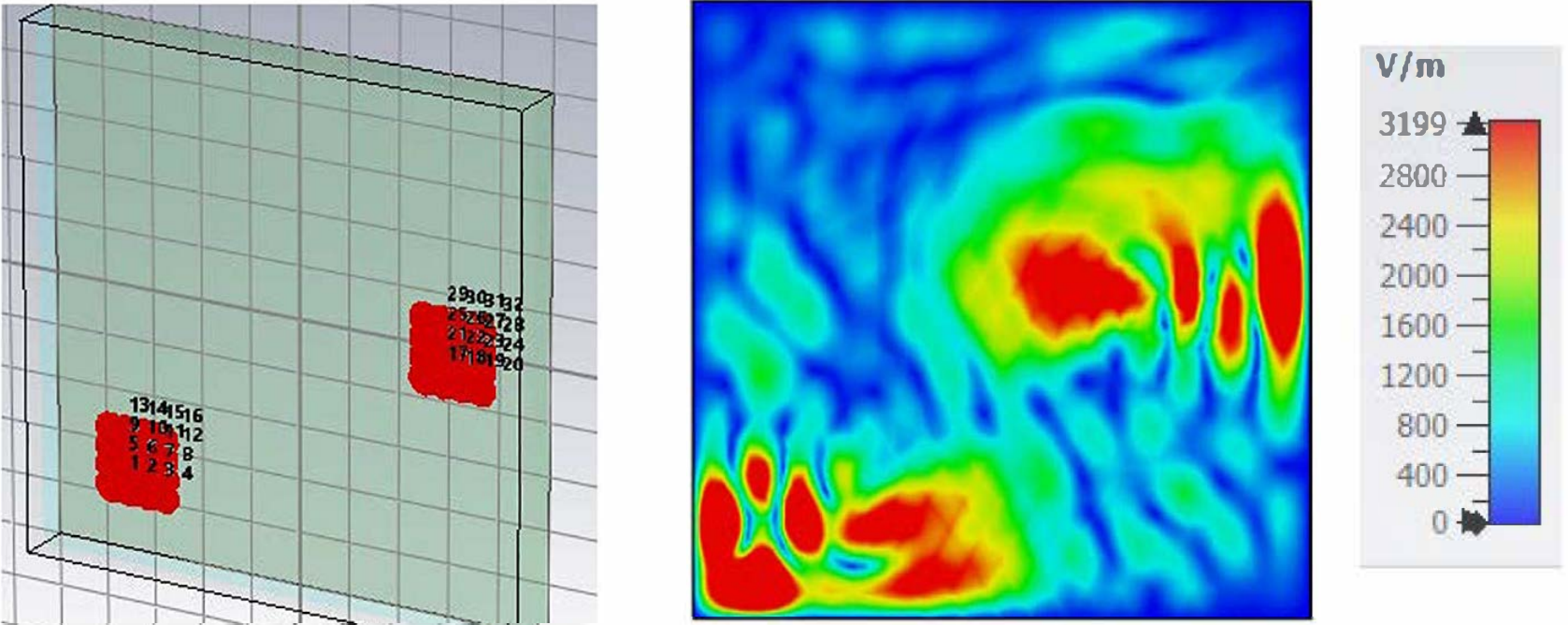}\vspace{-0.2cm}
 \caption{Array positions and field distribution.}
\end{subfigure}
\begin{subfigure}[b]{0.95\columnwidth}
 \centering
 \includegraphics[width=0.95\columnwidth]{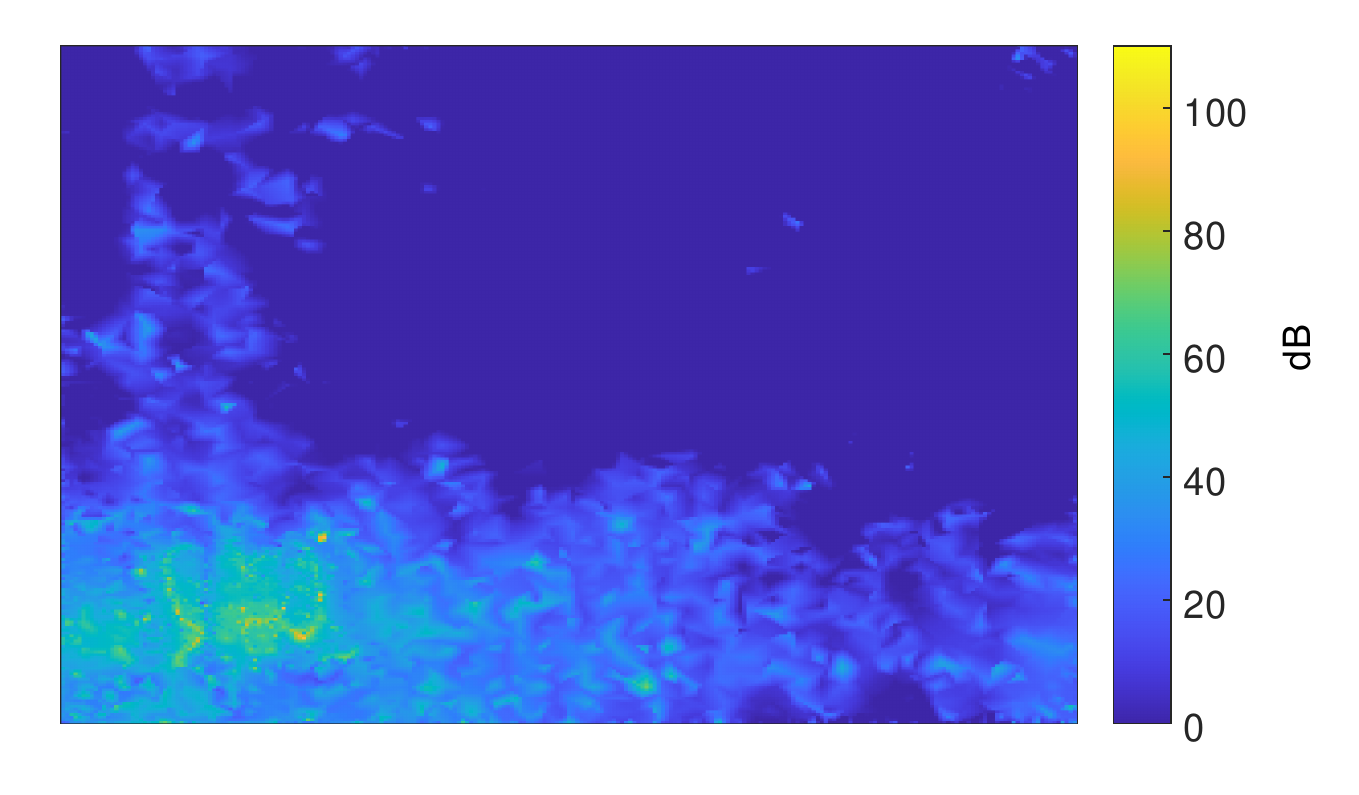}\vspace{-0.2cm}
 \caption{Signal-to-Interference Ratio from the bottom-left array.}
\end{subfigure}
\vspace{0.2cm}
\caption{Evaluation of an array system with nearby and potentially interferent arrays compatible with three spatial channels.}
  \label{fig:lateral_array}
\end{figure*}

\subsection{Scaling the Number of Channels} \label{sec:channels}
Next, we simulate the case of the arrays being placed closer to each other to see whether the scenario is compatible with more than two parallel spatial channels. In the new scenario, instead of two arrays on diagonal opposite corners of the chip, there is an array on the bottom left corner (array 1) and the other placed laterally and closer to the first one, as shown in~\figref{fig:lateral_array}. 

In this configuration, also two horizontal parallel channels seem plausible, yet with the interference taking a slightly different shape than in previous evaluations. Here, we see that array 1 (bottom-left corner) also produces interference along the Y-axis which hinders the error-free transmission from array 2. However, the achievement of good SIR even if the arrays are now closer is of high relevance in this scenario, because it could enable the integration of a third spatial channel in this hypothetical scenario.

\section{Conclusion}
\label{sec:conclusion}
In this paper, we made an analysis of the integration of a monopole antennas in an enclosed package to achieve concurrent multicast channels, which can be reconfigured by changing the excitation phases of the antenna elements. Understanding that it is not possible to direct and control the electromagnetic field by changing the phase of a single monopole, we considered antenna arrays. A review of the coupling issue among nearby antennas was performed to understand the tradeoff between undesired element coupling and compactness of the array. The first take away of our research is the confirmation of our ability to create, direct and somehow control the field distribution inside the chip with compact arrays of 1.18 mm\textsuperscript{2} at 60 GHz. Albeit considering a few idealities, especially in the phase shift of the excitation, we have shown that two and even three simultaneously radiating channels can be created with good values of SIR (beyond 20 dB). We approach to the possibility of scaling in frequency and in number of channel with some success, which opens the way towards spatial multiplexing and frequency diversity of the channels in on-chip environments. 

\section*{Acknowledgment}
Authors acknowledge support from the European Union’s Horizon 2020 research and innovation program, grant agreement 863337.

\bibliographystyle{ACM-Reference-Format}
\bibliography{bib2}

\appendix


\end{document}